\documentclass[sigconf]{acmart}

\usepackage{booktabs} % For formal tables
\usepackage[ruled]{algorithm2e} % For algorithms
\usepackage{graphicx}
\graphicspath{ {images/} }
\usepackage{amssymb}
\usepackage{amsmath}
\usepackage{commath}
\usepackage{url}

\SetAlFnt{\small}
\SetAlCapFnt{\small}
\SetAlCapNameFnt{\small}
\SetAlCapHSkip{0pt}
\IncMargin{-\parindent}

\setcopyright{acmcopyright}
\acmConference[SIGGRAPH Asia 2017]{SIGGRPAH Asia 2017}{November 2017}{Bangkok, Thailand} 
\acmYear{2017}
\copyrightyear{2017}
\acmPrice{15.00}

\setcitestyle{square}

\begin{document}
% Title portion
\title{Motion Style Extraction Based on Sparse Coding Decomposition}

\author{Xuan Thanh Nguyen}
\orcid{}
\affiliation{%
  \institution{NCCA, Bournemouth University}
  \streetaddress{}
  \city{} 
  \state{United Kingdom} 
  \postcode{}
}
\email{txnguyen@bournemouth.ac.uk}

\author{Thanh Ha Le}
\orcid{}
\affiliation{%
  \institution{HMI lab, UET-VNU}
  \streetaddress{}
  \city{Hanoi} 
  \state{Vietnam} 
  \postcode{}
}
\email{lthavnu@vnu.edu.vn}

\author{Hongchuan Yu}
\affiliation{%
  \institution{NCCA, Bournemouth University}
  \streetaddress{}
  \city{} 
  \state{United Kingdom} 
  \postcode{}
}
\email{hyu@bournemouth.ac.uk}

\renewcommand{\shortauthors}{X. T. Nguyen et al.}

\begin{abstract}
We present a sparse coding-based framework for motion style decomposition and synthesis. Dynamic Time Warping is firstly used to synchronized input motions in the time domain as a pre-processing step. A sparse coding-based decomposition has been proposed, we also introduce the idea of core component and basic motion. Decomposed motions are then combined, transfer to synthesize new motions. Lastly, we develop limb length constraint as a post-processing step to remove distortion skeletons. Our framework has the advantage of less time-consuming, no manual alignment and large dataset requirement. As a result, our experiments show smooth and natural synthesized motion.
\end{abstract}


\thanks{NCCA: National Centre of Computer Animation, Bournemouth University;
HMI Lab, UET-VNU: Human-Machine Interaction Laboratory, University of Engineering and Technology, Vietnam National University Hanoi;
This work was supported by H2020-MSCA-RISE ANIAGE and Multimedia Application Tools for Intangible Cultural Heritage Conservation and Promotion (DTDL.CN-34/16) funded by Vietnam Ministry of Science and Technology.}

\maketitle

%SEC 1------------------------------------------------------------------------------------------------------------------------
\section{Introduction} \label{sec:1}

In modern times, with the rapid development of computing technology, the realistic synthesis of human motion plays an important role in animation, film production, and digital entertainment. \cite{minchai12} A major goal of human performance capture and animation is to reconstruct and simulate realistic human behaviors, which benefits many downstream applications. For example, this will help enhance the sense of immersion for virtual reality. However, it is a challenging problem, because human performance includes diverse poses and complex motions. Motion related processing is expensive and time-consuming, therefore an alternative approach considers reusing captured motion data to synthesize new motions by analyzing existing motions, to satisfy diverse environmental constraints.

Some methods have been proposed, they mainly use linear models on motion datasets of walking, running \cite{hsu05} \cite{xia15} or recently one machine learning-based method using a convolutional neural network \cite{holden16} . However, those approaches are typically time-consuming, require large training dataset of various styles and manual alignment. Besides, foot-skating and abnormal synthesized motions are other problems. 

Because capturing high-quality motion time-consuming, expensive, requiring professional marker system and actors which are not always available, it is necessary to reproduce motion by analyzing, synthesizing available motions for various applications. In this paper, we aim to build a framework to extract human motion features, named motion styles to transfer to other models by synthesis or other techniques. Motion styles can be used to the reconstructed original motion, as well as synthesizing new realistic motions. 

Our main contributions are:
\begin{itemize}
\item Propose a style extraction framework using motion decomposition;
\item Real-time motion synthesis without the requirement of large dataset;
\item Overcome foot-skating using Dynamic Time Warping.
\end{itemize}

The remainder of this paper is organized as follows: Section 2 gives a short literature review of state-of-the-art human motion studies. Section 3 describes our proposed framework of human motion synthesis, dynamic time warping and limb length constraints. Experimental results and observations are given in Section 4. Finally, in Section 5, we briefly conclude our method and contributions. 

%SEC 2------------------------------------------------------------------------------------------------------------------------
\section{Related Work}\label{sec:2}

In order to capture full body motion, there are two kinds of motion capture systems, i.e. marker based and marker-less motion capture systems. Traditional marker-less MoCap methods usually extract 3D human skeleton data from 2D image sequences \cite{deepcut16} \cite{2d3d15}. Recently, there are many cheap 3D scanners (e.g. MS Kinect) available to be involved in it. Although this kind of systems is low cost, they are usually imprecise and sensitive to the external environment. On the other hand, expensive MoCap systems capture 3D human skeleton precisely, but wearing marker suit somehow reduces character's expression and naturalness. Reproducing existing high-quality motions is necessary to keep synthesized motion realistic and low-cost. 

State-of-the-art approaches attempted to transfer style including emotion from one motion to another. Hsu et al. and Xia et al. \cite{hsu05} \cite{xia15} introduces linear time invariant models, a mixture of autoregressive models and a search scheme on huge motion dataset to find appropriate style candidate frame by frame. Holden et al. \cite{holden16} applied convolutional neural network to map motion sequence to a motion manifold of hidden unit, then perform an optimization to synthesize motion style. From another point of view, Yumer et al. \cite{fft16} converts conventional motion data into the frequency domain, then do synthesize based on the difference among styles. Some other studies focus on emotion recognition \cite{piana16} \cite{wang16}. Those methods typically perform searches or training which requires a huge motion dataset of all independent actions and styles. Another drawback is requiring manual alignment and labeling.

In contrast, we propose a framework which decomposes full body motion into components, named motion style, using matrix factorization and sparse coding. We further show a synthesis scheme and a post-processing of limb length constraints to remove distortion on the synthesized motion. As a result, our framework can decompose, synthesize character motion smoothly without any manual alignment or large database. We also successfully clean distortion from the synthesized animation.
%SEC 3------------------------------------------------------------------------------------------------------------------------
\section{Proposed Method}
\label{sec:3}

%SUBSEC ---------------------------------------------------------
\subsection{System Overview}
\begin{figure}[!ht]
\centering
\includegraphics[width=0.99\columnwidth]{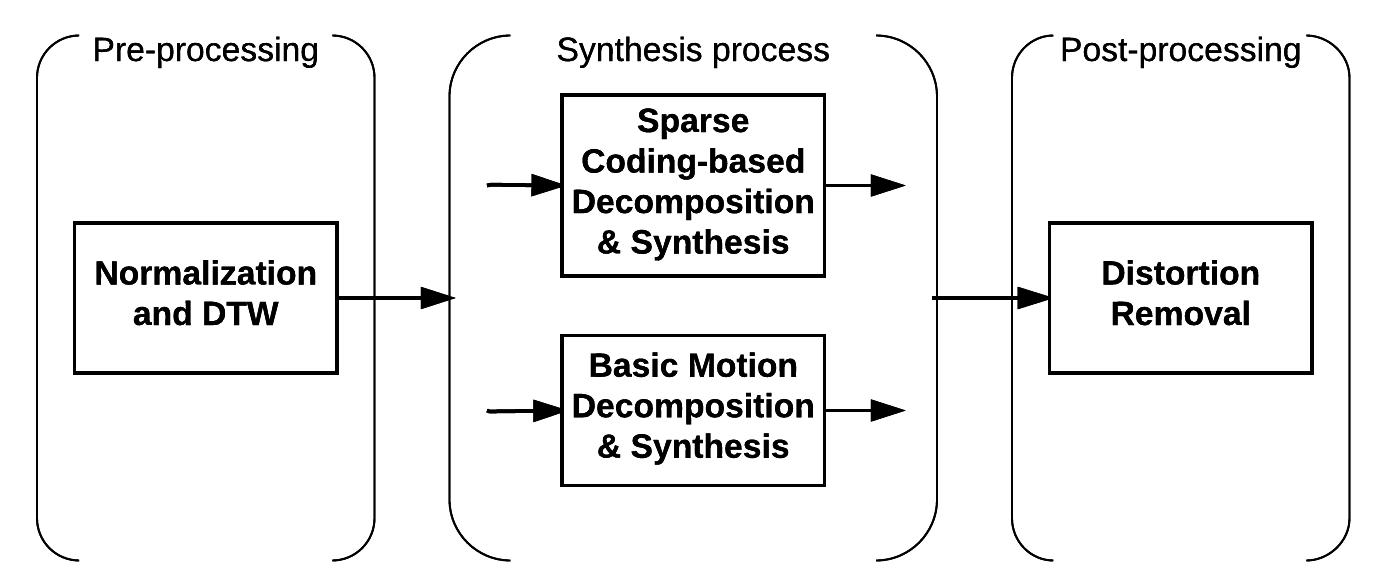}
\caption{Overview of our framework.}
\label{fig:overview}
\end{figure}

Our framework has three main stages as shown in Figure \ref{fig:overview}. In pre-processing, motion sequences are retargeted, normalized and synchronized by Dynamic Time Warping (DTW). Subsequently, synthesis-process extracts motion features/styles using sparse coding-based decomposition. Motion features and styles are combined to synthesized new motions. Lastly, post-processing applies limb length constraints to clean skeleton distortion.

%SUBSEC ---------------------------------------------------------
\subsection{Motion Data Acquisition}\label{sec:data_normalize}
We use a subset of the CMU Motion Capture Database which was introduced by Carnegie Mellon University \cite{cmu}. In our research scope, we only select walking motions with different styles and moods (e.g.neutral, angry, happy, sad, old). The subset is then normalized in several steps. Firstly, we retarget subset motions into a unique skeleton. As a consequence, all retargeted motion have the same bone lengths. Secondly, we convert original rotational motion into 3D positional data, then selecting 20 major joints and sub-sampling data to frame-rate of 30 fps (originally, 120 fps). In order to have zero mean and standard deviation of one, motion data is subtracted by mean pose and divided by standard deviation.

%SUBSEC ---------------------------------------------------------
\subsection{Dynamic Time Warping for Motion}
\begin{figure}[!ht]
\centering
\includegraphics[width=0.99\columnwidth]{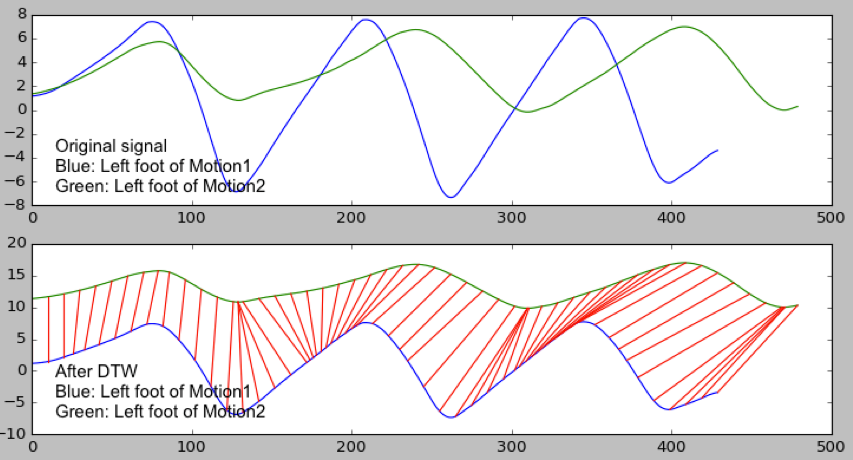}
\caption{DTW visualization of two actors' left feet movement.}
\label{fig:dtw}
\end{figure}

Dynamic Time Warping (DTW) is an algorithm to measure the similarity between signals in time domain \cite{dtw78,dtw95}. For motion, we target to use DTW to synchronize two arbitrary motions, even actors are walking or running \cite{dtw16}.  We time-warp two motions by time-warp their corresponding left feet signals. As can be seen from Figure \ref{fig:dtw}, upper-part indicates two original feet signals; lower-part shows synchronized/warped signals; red lines exactly show the way to map each data series of first motion with appropriate data series in second motion. 

%SUBSUBSEC -------------------------------------------------------------------------------------------
\subsection{Sparse Coding}\label{sub:sec:sparsecoding}
Sparse coding cost function.
$$ \min_{x_j \in R^n}{\frac{1}{2}\norm{y - D \cdot x_j}}^2, s.t. \norm{x_j}_0 < k, j = 1\ldots m$$
The usual iterative scheme, 
$$ x_j^{k+1} = x_j^k - (D^\top D)^{-1} D^\top (D \cdot x_j - y)$$
Herein, rewrite it as,
$$ x_j^{k+1} = x_j^k - \frac{2}{norm(D^\top D)} D^\top (D \cdot x_j - y)$$
i.e. to simply computation, select the smallest step-length.
Then, orthogonally projecting the resulting x onto sparse solution space yields,
$$x_i =  
\begin{cases} 
	x_i &	\text{if } \abs{x_i} \geq \bar{x}_k\\
	0 &	otherwise
\end{cases} $$
Where $i = 1, \ldots, n$ and $\bar{x}$ usually depends on applications. This concludes the sparse code $x$. In terms of motion decomposition application $y = D \cdot x$, where $y$ denotes the original motion matrix, $D$ is the weight matrix and $x$ denotes the components. The choice of $\bar{x}$ sparsity determines how many joints are employed in the current recovered motion. 

%SUBSUBSEC -------------------------------------------------------------------------------------------
\subsection{Sparse Coding-based Decomposition}\label{sub:sec:optimization}
The input data to the algorithm consists of a $F$ frames motion. Each frame consists of $N$ joint positions $v_i^{(f)}$. The skeleton structure is shared for all frames. We arrange a single animation matrix $X \in \mathbb{R}^{F \times 3N}$ by stacking the joints of all frames row-wisely as follows.
\begin{equation}
\mathbf {X} ={
\begin{bmatrix}
(v_{1}^{(1)})^\top&(v_{2}^{(1)})^\top&\cdots &(v_{N}^{(1)})^\top\\
(v_{1}^{(2)})^\top&(v_{2}^{(2)})^\top&\cdots &(v_{N}^{(2)})^\top\\
\vdots &\vdots &\ddots &\vdots \\
(v_{1}^{(F)})^\top&(v_{2}^{(F)})^\top&\cdots &(v_{N}^{(F)})^\top
\end{bmatrix}}
\end{equation}

We are looking for an appropriate matrix factorization of $X$ into $K$ deformation components $C \in \mathbb{R}^{K\times3N}$ with weights  $W \in \mathbb{R}^{F\times K}$.

\begin{equation} \label{eq1}
X = W C
\end{equation}

In order to solve Eq. \ref{eq1} with sparse components, we add regularization term as below. Then, the matrix factorization can be formulated as an optimization problem.

\begin{equation}\label{eq3}
\underset{W,C}{\arg \min}{\norm{X - WC}_F^2} + \Omega(C)  \text{	} s.t. V(W)
\end{equation}

\textbf{Optimization}
In order to solve the non-convexity optimization problem of Eq. \ref{eq3}, we iteratively fix weights $W$ to update components $C$, then fix $C$ to estimate $W$ [ref Mairal09]. Our optimization process has three main stages as follows:

\paragraph{Initialization}
Given a normalized motion matrix $X$, we initialize $K$ pairs of (component $C_i$, weight $W_i$) iteratively. Each looping, variance matrix of $X$ is defined as horizontal summation of normalized motion. $$Var(X) = \underset{axis=1}{Sum}(X^2) $$ where $axis = 1$ means horizontal summation, $axis=0$ means summation vertically. By looking at $Var(X)$, vertex which has maximal variance value will be chosen. Factorizing motion of chosen vertex into component $C_i$ and weight $W_i$. Then, reconstructing motion matrix, updating residual matrix. $$X' = W_i C_i$$  $$X \gets X-X'.$$ This process continues until we obtain $K$ pairs of components and weights. The chosen components and weights is definitely not optimal solution, but the components were initialized sparsely.

\paragraph{Weights optimization}
Given the motion matrix $X$ and initial components C, problem of Eq. \ref{eq3} respect to the weights $W$ is a constrained linear least squares problem. The problem is separable, then constraints act on the weight vector $W_{:, k} $of each component separately.
$$\underset{W_{:.k} \in V}{\arg \min}\norm{X - WC}_F^2 = P_v({\frac{( (X-WC)+W_{:,k}C_k)C_k}{{C_k}^\top C_k}}) $$
where $P_v$ performs a projection of the weights onto the constraint set, depending on which constraint is used. 

\paragraph{Components optimization}
Given the motion matrix $X$ and fixed weight $W$, we find best sparse components $C$ using $L2/L0$ sparse coding which is mentioned in subsection \ref{sub:sec:sparsecoding}. In terms of sparse coding, we set sparsity threshold to control percentage of non-zeros elements in components $C$. To this end, we obtain optimized sparse components $C$ and weights $W$ for the motion data $X$.

%SUBSUBSEC -------------------------------------------------------------------------------------------
\subsection{Basic Motion Decomposition}
Base on Sparse Coding-based decomposition, we propose an idea of basic motion decomposition which use a set of basic motion to represent original motion data. Suppose that the motion data X can be decomposed by,
$$X_{F\times N} = W_{F\times K} C_{K\times N}$$
Apply sparse coding to the component dimension $K$ with $L2/L0$ norm along the 3D coordinate dimension $N$. How to use sparse coding and optimize decomposition process were described details in sub-section \ref{sub:sec:optimization}. 

Now, extracting a basic motion from $X$ by set sparsity coefficient $fi$, update components $C_{fi}$, and update residual motion.
$$X_{fi} = W{fi}C_{fi} $$
$$X \gets X - X_{fi}  $$
The idea is not only obtaining one basic motion, but representing motion by a chain of $m$ basic motions. Then we extract the group of $m$ basic motions by iteratively updating sparsity, components and residual motion. The sparsity coefficient may be either same or different depending on applications. To this end, a motion data $X$ is decomposed into $m$ basic motion.
$$ X= \sum_{i=1}^m{X_{fi}} = \sum_{i=1}^m{W_{fi}C_{fi}} $$

%SUBSEC ---------------------------------------------------------
\subsection{Motion Synthesis} \label{sec:syn_core}
In this section, we introduce how to synthesize motions using sparse coding decomposed components and basic motion.

\paragraph{Components SVD}
For an input motion data sequence $X$, we take decomposition by method \ref{sub:sec:optimization} and yield $X = WC$, where $C$ contains $k$ components. Then we take Singular Value Decomposition (SVD) on $C$. 
\begin{equation}
C = (U\Sigma) V^\top =  K V^\top.
\end{equation}
where $K = U\Sigma$ named core component.

\paragraph{Components synthesis}
We synthesize two motion sequences $X^1$ and $X^2$ using their corresponding core components. 

\begin{equation}
\begin{split}
X^1 &= W^1 K^1 V^{1\top},\\
X^2 &= W^2 K^2 V^{1\top},\\
X_{syn} &= W^1( (1-\alpha)I K^1 + \alpha I K^2) V^{1\top},\\
\alpha \in [0,1] &, I \text{ denotes identity matrix}.
\end{split}
\end{equation}

Number of component $k$ and alpha are two tunable parameters that may vary on applications. The bigger number of $k$, the better motion decomposition and reconstruction. Alpha decides how many percent of input motions contribute to synthesized motion.

\paragraph{Basic motion SVD}
Similar to Components SVD, given a basic motion, we apply Singular Value Decomposition on a basic motion as below,
\begin{equation}
\begin{split}
X_{fi} &= W_{fi}C_{fi} = W_{fi}(U_{fi}\Sigma_{fi} V_{fi}^\top) \\
	&= W_{fi}(U_{fi}\Sigma_{fi}) V_{fi}^\top = W_{fi} K_{fi} V_{fi}^\top\\
\text{where  }K_{fi} &= U_{fi}\Sigma_{fi}
\end{split}
\end{equation}
We called $K_{fi}$ core component.

\paragraph{Basic motion synthesis}
The idea of emotion synthesis is combining core component $K_{fi}$ of one motion to another. Given two motions, we replace core component $K^1_{fi}$ of $X1$ by core component $K^2_{fi}$ of $X^2$ as below,
\begin{equation}
\begin{split}
X^1 &= \sum_{i=1}^m{W^1_{fi}K^1_{fi}} V_{fi}^{1\top}\\
X^2 &= \sum_{i=1}^m{W^2_{fi}K^2_{fi}} V_{fi}^{2\top}\\
X_{syn} &= W^1_{f1} K^1_{f1} V_{f1}^{1\top} +\ldots + W^1_{fi} \mathbf{K^2_{fi}} V_{fi}^{1\top} + \ldots\\
&+ W^1_{fm} K^1_{fm} V_{fm}^{1\top}\\
\end{split}
\end{equation}

%SUBSEC ---------------------------------------------------------
\subsection{Motion Post-processing}
\begin{figure}[!ht]
\centering
\includegraphics[width=0.95\columnwidth]{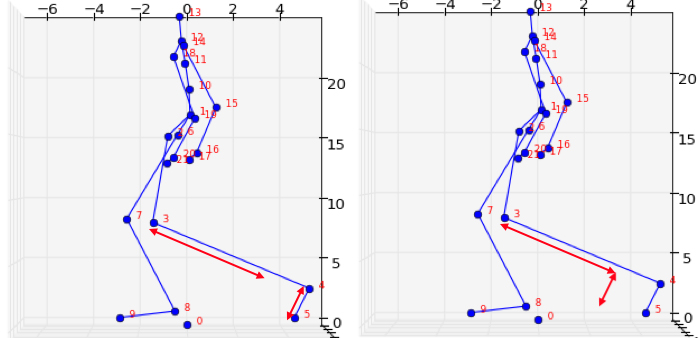}
\caption{Distortion removal process.}
\label{fig:dis_1}
\end{figure}

\begin{figure}[!ht]
\centering
\includegraphics[width=\columnwidth]{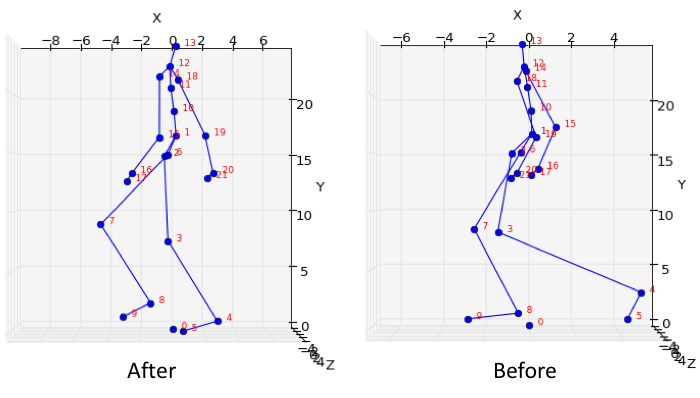}
\caption{Skeleton with and without distortion removal.}
\label{fig:dis_2}
\end{figure}

Animating human motion usually causes distortion and foot-skate on the generated motion. This paper proposed a simple post-processing step to remove those abnormal artifacts. It is obvious that human limbs length is biologically unchanged in a motion sequence. In order to remove distortion, we firstly calculate original limbs length, then applying those limbs length on the animated motion. By this way, corrected motion limbs length is preserved as input motion. Figure \ref{fig:dis_1} visualizes the distortion removal process. Figure \ref{fig:dis_2} shows abnormal skeleton before and after applying our proposed distortion clean up. 

%SEC 4------------------------------------------------------------------------------------------------------------------------
\section{Experimental Results}
\label{sec:4}

%SUBSEC ---------------------------------------------------------
\subsection{Motion Synthesis using Sparse Coding Components}
\begin{figure}[!ht]
\centering
\includegraphics[width=0.9\columnwidth]{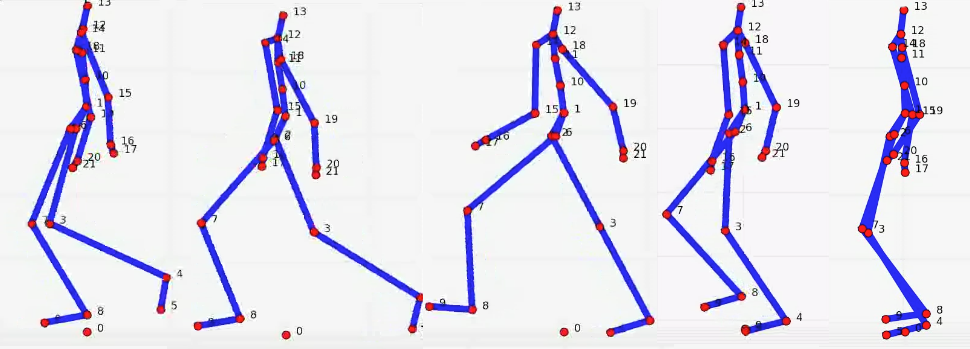}
\caption{Normal walking motion.}
\label{fig:exp_1_1}
\end{figure}

\begin{figure}[!ht]
\centering
\includegraphics[width=0.9\columnwidth]{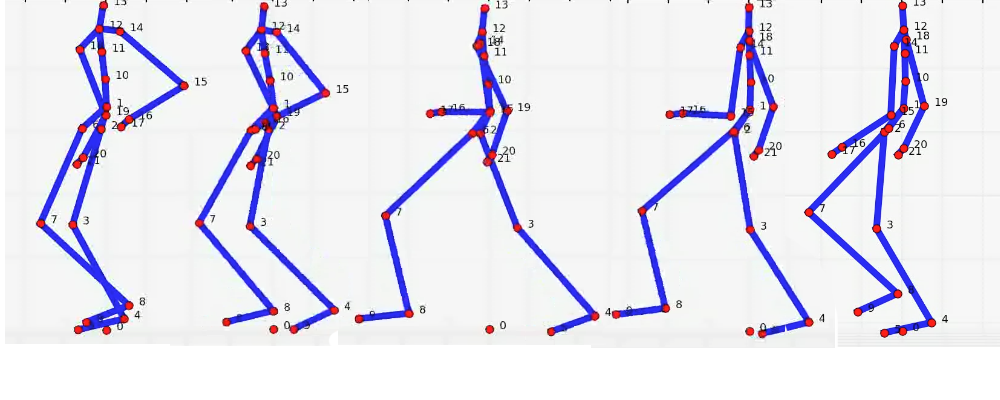}
\caption{One-leg-hurt walking motion.}
\label{fig:exp_1_2}
\end{figure}

\begin{figure}[!ht]
\centering
\includegraphics[width=0.9\columnwidth]{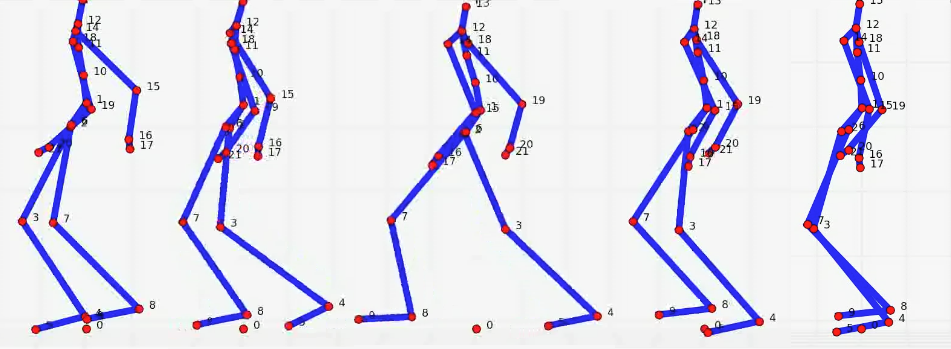}
\caption{Synthesize motion of normal and one-leg-hurt sequences.}
\label{fig:exp_1_3}
\end{figure}

We demonstrate our framework on a variety of captured walking motion sequences of CMU Motion Capture Database \cite{cmu} (20 motions sequences of different style/mood have been chosen for testing). This experiment attempts to synthesize two walking motion sequences of different moods using sparse coding-based decomposition. Two input walking motions are firstly synchronized by Dynamic Time Warping technique. Subsequently, we use sparse coding decomposition to extract components $C$ and core components $K$ from warped motions. Base on estimated components, we combine two motion using formula described in section \ref{sec:syn_core}. We can control the contribution of each input motions by adjusting the number of component $k$ and coefficient $\alpha$. Finally, limb length constraints remove skeleton distortion to keep synthesized motions natural. Figure \ref{fig:exp_1_1}, \ref{fig:exp_1_2}, \ref{fig:exp_1_3} show the combination (sparsity $f = 0.1; k = 10; \alpha = 0.5$) of normal walking motion and one-leg-hurt walking motion produce synthesized motion. Demo videos for other sparsity ($f \in [0,1]$) synthesis are available here \footnote{\url{https://goo.gl/FKxWkz}}

%SUBSEC ---------------------------------------------------------
\subsection{Motion Synthesis using Basic Motion}
\begin{figure}[!ht]
\centering
\includegraphics[width=0.9\columnwidth]{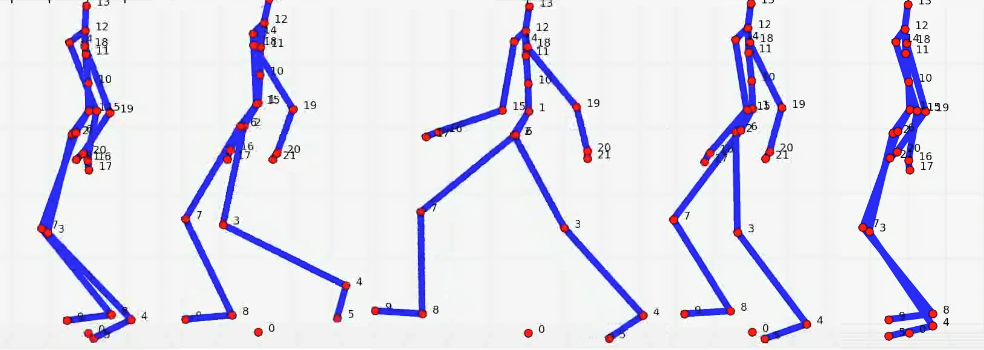}
\caption{Exchanged basic motion of normal and one-leg-hurt sequences.}
\label{fig:exp_2_3}
\end{figure}

Similarly, this experiment processes three steps synthesis of pre-processing, decomposition and post-processing. While the previous experiment represents input motions using sparse coding core components, this experiment represents input motions by a chain of basic motions. Afterwards, we synthesize two motions by exchanging core components of one selected basic motion, formula described in section \ref{sec:syn_core}. Figure \ref{fig:exp_2_3} describes synthesized sequence by exchanging their corresponding basic motion core components between normal motion and one-leg-hurt walking motion.

%SEC 5------------------------------------------------------------------------------------------------------------------------
\section{Conclusion}
\label{sec:5}
In this paper, we develop a sparse coding-based style decomposition method to reproduce, synthesize available motion. Unlike existing methods, our framework does not require large motion database and manual alignment. In order to synchronize motion sequences, we use dynamic time warping as a pre-processing. Besides, a limb length constraint has been proposed to eliminate abnormal synthesized motions. Consequently, core component experiment presents the result of synthesizing two different style motions, controlled by coefficient $\alpha$. Basic motion experiment shows how to transfer a core component from one motion to another. Experiments clearly show synthesized motion are realistic and smooth.

\bibliographystyle{ACM-Reference-Format}
\bibliography{template} 
\end{document}